\input{epsfig}
\documentstyle[prd,aps,twocolumn]{revtex}
\catcode`\@=11

\def\maketitle2{\par % Uses \twocolumn[\@maketitle2].
\begingroup
\let\cite\@bylinecite
\def\thefootnote{\fnsymbol{footnote}}%
\twocolumn[\@maketitle2\vskip2pc]%
\thispagestyle{plain}\@thanks
\endgroup
\def\thefootnote{\arabic{footnote}}%
\setcounter{footnote}{0}%
\let\maketitle2\relax \let\@maketitle2\relax
\let\@thanks\relax \let\@authoraddress\relax \let\@title\relax
\let\@date\relax \let\thanks\relax \let\@abstract\relax 
\let\@pacs\relax}

\def\abstract#1{\gdef\@abstract{{\par % Store abstract text. 
\bgroup
\ifdim\prevdepth=-1000pt \prevdepth0pt\fi
\hsize\columnwidth
\dimen0=-\prevdepth \advance\dimen0 by17.5pt \nointerlineskip
\small\vrule width 0pt height\dimen0 \relax}{~~}#1\egroup}}

\def\pacs#1{\gdef\@pacs{{\par % Store PACS numbers as \@pacs.
\bgroup
\hsize\columnwidth \parindent0pt
\ifdim\prevdepth=-1000pt \prevdepth0pt\fi
\dimen0=-\prevdepth \advance\dimen0 by20pt\nointerlineskip
\egroup} PACS numbers:~#1}}

\def\@maketitle2{% Puts \@abstract and \@pacs in a {list}.
\@preprint
\@title
\ifdim\prevdepth=-1000pt \prevdepth0pt\fi
\@authoraddress
\@date
\begin{list}{}{\leftmargin=0.10753\textwidth \rightmargin=\leftmargin
\itemsep=1pc\partopsep=-1pc}
\item\@abstract
\item\@pacs
\end{list}
}

\catcode`\@=12

\begin{document}
\draft
\preprint{LA-UR-97-4742}
\title{Statistical Mechanics of Double sinh-Gordon Kinks} 
\author{Salman Habib$^{1\ast}$, Avinash Khare$^{2}$, and Avadh
Saxena$^{3}$}  
\address{{$^1$}T-8, Theoretical Division, MS B285, Los Alamos National
Laboratory, Los Alamos, New Mexico 87545} 
\address{$^2$Institute of Physics, Sachivalaya Marg, Bhubaneswar 751
005, India} 
\address{{$^3$}T-11, Theoretical Division, MS B262, Los Alamos National
Laboratory, Los Alamos, New Mexico 87545} 

\date{\today}
\abstract
{We study the classical thermodynamics of the double sinh-Gordon (DSHG)
theory in 1+1 dimensions. This model theory has a double well
potential $V(\phi)=(\zeta\cosh 2\phi-n)^2$ when $n>\zeta$, thus
allowing for the existence of kinks and antikinks. Though it is
nonintegrable, the DSHG model is remarkably amenable to
analysis. Below we obtain exact single kink and kink lattice solutions
as well as the asymptotic kink-antikink interaction. In the continuum
limit, finding the classical partition function is equivalent to
solving for the ground state of a Schr\"odinger-like equation obtained
via the transfer integral method. For the DSHG model, this equation
turns out to be quasi-exactly solvable. We exploit this property to
obtain exact energy eigenvalues and wavefunctions for several
temperatures both above and below the symmetry breaking transition
temperature (provided $n=1,2,\cdots,6$). The availability of exact
results provides an excellent testing ground for large scale Langevin
simulations. The probability distribution function (PDF) calculated
from Langevin dynamics is found to be in striking agreement with the
exact PDF obtained from the ground state wavefunction. This validation
points to the utility of a PDF-based computation of thermodynamics
utilizing Langevin methods. In addition to the PDF, field-field and
field fluctuation correlation functions were computed and also found
to be in excellent agreement with the exact results.}

\pacs{05.20.-y, 11.10.-z, 63.75.+z, 64.60.Cn}    

\maketitle2
\narrowtext

\section{Introduction}

The double-well $\phi^4$ model in $1+1$ dimensions has been
extensively studied in the context of symmetry breaking transitions.
As is well-known, the equilibrium classical statistical mechanics of a
$1+1$ dimensional field theory reduces to a time-independent quantum
mechanics problem via the transfer integral method
\cite{ssf,krum}. However, the Schr\"odinger equation with a $\phi^4$
potential does not have any known exact solutions. To overcome this
problem, we turn to another double-well system that is quasi-exactly
solvable (QES), where by a QES problem we mean one in which the exact
{\em partial} diagonalization of the Hamiltonian may be carried out
\cite{qes}. Knowing just the low-lying eigenvalues and eigenfunctions
is sufficient to obtain almost complete information regarding the
classical thermodynamics of the field theoretic system. As will be
seen below the QES property holds only at a discrete set of
temperatures. Our strategy is to tune system parameters so as to
straddle an interesting region in temperature space which, for this
paper, we have taken to be the region around the short-range order
(``kink'') transition point.

An important motivation for exact results is validation of numerical
techniques: recent advances in large scale computation have made it
possible to calculate quantities such as the probability distribution
function (PDF) using Langevin dynamics to very high accuracy
\cite{sh,khs}. The PDF is the probability distribution of field values
averaged over the total system volume.  Alternatively, it is equal to
the square of the ground state wavefunction of the transfer operator
at a given temperature.  All thermodynamic information can be shown to
reside in the PDF \cite{pdf,PDF}.  Calibration against the PDF
calculated from the exact ground state wavefunction is an essential
guide when performing error and convergence analysis of the Langevin
equation \cite{shgl}.

The double-well QES system that we focus on in this paper is specified
by the double sinh-Gordon potential $V(\phi) = (\zeta\cosh 2\phi
-n)^2$, where $\zeta$ is a positive parameter and $n$ is a positive
integer \cite{khs,razavy,khare,dsinh,dinda,kofane,konwent}. This
potential is the hyperbolic analog of the double sine-Gordon system
\cite{bullough}.  Similar potentials arise in the context of the
quantum theory of molecules ({\em e.g.}, a homonuclear diatomic
molecule), wave motion describing the normal modes of vibration of a
stretched membrane of variable density \cite{ince}, and as the
solution of a Fokker-Planck equation \cite{razavy}.  Note that the
hyperbolic analog of the sine-Gordon equation is a single well
potential (sinh-Gordon \cite{sinh}) and thus uninteresting from the
soliton statistical mechanics perspective.

Below we find exact solutions for the ground and a few excited state
energy eigenvalues and wavefunctions for certain temperatures both
above and below the short-range transition temperature. This allows
analytic calculation of the PDF and correlation functions ($C_1$ and
$C_2$) and direct comparison with those calculated from a Langevin
simulation. The analytic kink profile can also be compared with the
corresponding field configuration in the simulation. We find striking
agreement between the exact results and simulations. The
proven high quality of the simulations provides an important
validation for PDF-based thermodynamics implemented via Langevin
methods.

Before proceeding to a more detailed view of the DSHG model we
contrast its main features with those of other models that have been
used previously for analytic studies. The Schr\"odinger equation with
a double-quadratic potential $V(\phi) = \frac{1}{2}(|\phi| -1)^2$
\cite{trudel,currie} is in fact exactly solvable. However, the energy
eigenvalues are not known as simple functions but may be obtained only
as a solution of transcendental equations involving parabolic cylinder
functions.  Moreover, this potential is not a smooth function and has
a cusp at $\phi = 0$.  Other smoothly varying double-well potentials,
{\em e.g.}, the Manning potential
$V(\phi)=-A~\hbox{sech}^2(\phi/2\rho)+B~\hbox{sech}^4(\phi/2\rho)$
\cite{manning} and the double-Gaussian model
$V(\phi)=1/2(\phi/\rho)^2- \ln~\cosh(\phi\sigma/\rho^2)$,
\cite{dgaussian} do not allow exact or quasi-exact solvability of the
associated Schr\"odinger equation. We note that the double-Morse
potential, which arises in several physical contexts such as
hydrogen-bonded chains \cite{hbond}, is closely related to the DSHG
potential.

\section{The double sinh-Gordon model} 

We start with the continuum representation of a model Hamiltonian
describing a system capable of undergoing a displacive transition:
\begin{equation}
\label{1} 
H=\int\frac{dx}{l}\left[\frac{m}{2}\phi_t^2
+V(\phi)+\frac{mc_0^2}{2}\phi_x^2 \right], 
\end{equation} 
where $l$ is the lattice spacing, $m$ the mass of particles (ions) and
$c_0$ the velocity of low-amplitude sound waves (phonons) in the
associated discrete problem \cite{krum}.  Here the potential is
\begin{equation}
V_{DSHG}(\phi) = (\zeta\cosh 2\phi-n)^2~, \label{1a}
\end{equation} 
where $\zeta$ is a positive parameter. While the value of $n$ is not
restricted in principle, it has to be a positive integer for the QES
property to hold, along with $n>\zeta$ in order to have a double well.
The two minima occur at $\cosh2\phi_0 = n/\zeta$ with $V_{min}=0$ and
a local maximum at $\phi=0$ with $V(0)=(n-\zeta)^2$. We depict the
potential for a few values of $n$ and $\zeta$ in Fig.~1.

The DSHG potential written in the form (\ref{1a}) has all the generic
features of a double-well potential such as Landau-Ginzburg, but
allows for much greater analytic progress. Below we find exact
solutions for $(1)$ a kink, $(2)$ a kink lattice, $(3)$ phonon
dispersion, and $(4)$ the first few eigenvalues and eigenfunctions of
the transfer operator at certain temperatures both above and below the
transition, thereby allowing analytic calculation of the PDF and
correlation functions in the thermodynamic limit.

%PUT FIGURE 1 HERE:
\vspace{.5cm}
\centerline{\epsfig{figure=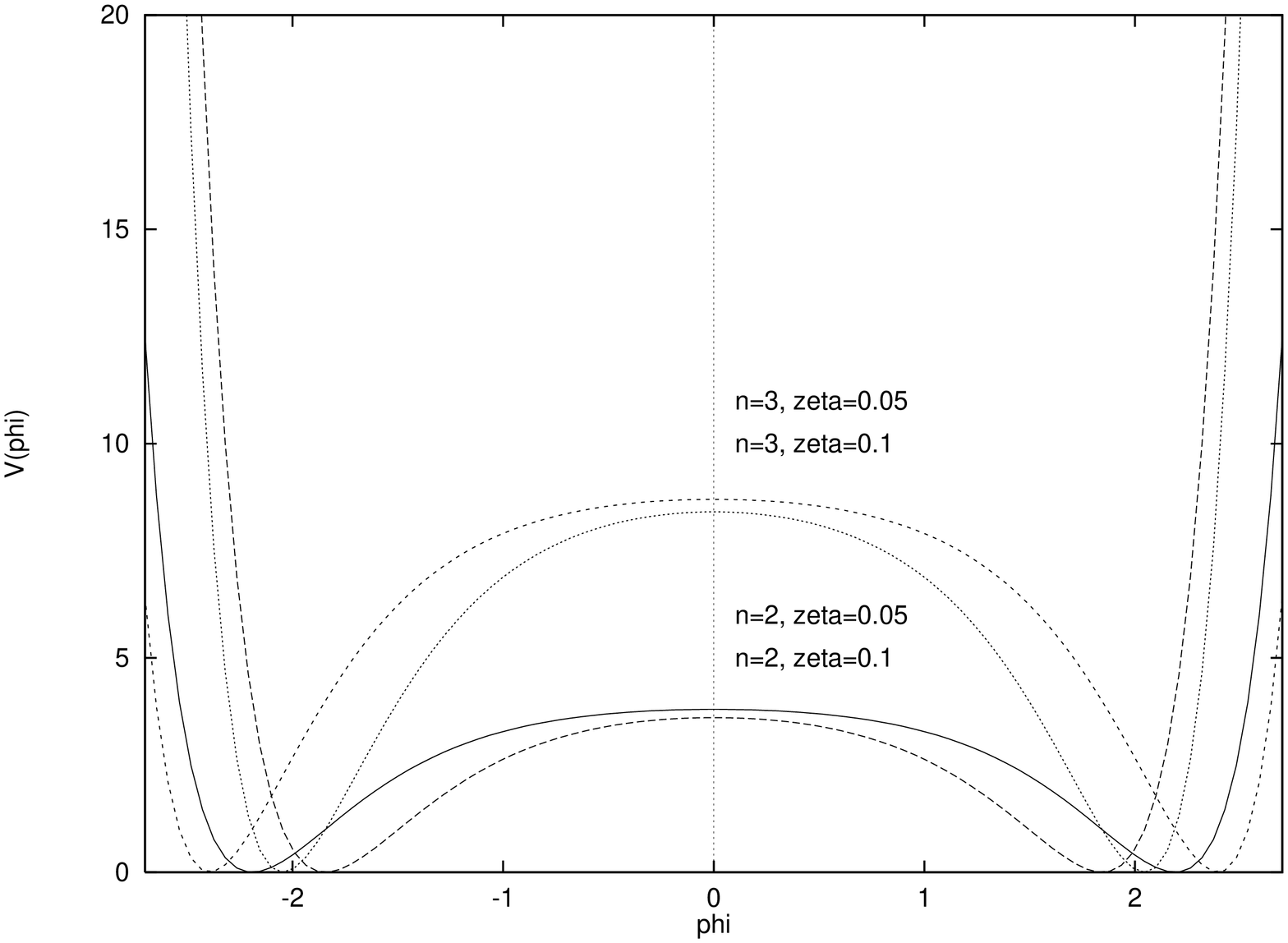,height=5.5cm,width=8cm,angle=-90}}
\vspace{.25cm}
\centerline{FIG. 1. {\small{The DSHG potential for $n=2,3$ and
$\zeta=0.05,0.1$.}}}
\vspace{.5cm}

\section{Kink and Kink Lattice Solutions} 

The single kink solution for the double sinh-Gordon potential is 
nothing but the finite energy solution of the equation of motion  
${1 \over 2}g\phi_x^2 (x) = V_{DSHG} (\phi)$.  The kink/antikink 
located at $x_0$ is given by \cite{khs,khare}  

\begin{eqnarray}\label{2} 
\phi(x) &=& \pm \tanh^{-1}\left(\sqrt{\frac{n-\zeta}{n+\zeta}}
\tanh\frac{x-x_0} {\xi}\right); \nonumber\\
\xi &=& \sqrt{\frac{g}{2(n^2-\zeta^2)}}; ~~~n > \zeta~.      
\end{eqnarray} 
Here $g = mc_0^2$ and $\tanh\phi_0=\sqrt{(n-\zeta)/(n+\zeta)}$. The
constant $g$ is often introduced in condensed matter treatments as a
phenomenological parameter and controls the kink size. Note that
instead of the potential (\ref{1a}) one could consider a more general
potential of the form $a(\zeta \cosh 2b\phi -n)^2$. It is easily seen
that the kink solution (\ref{2}) is unchanged except for the
replacement of $g$ by $g/ab^2$. For simplicity, throughout this paper
we take $a,b =1$ since solutions for the general case are trivially
obtained from here. Traveling kink solutions are obtained by boosting
to velocity $v$ via $x \rightarrow (1-v^2)^{-1/2} (x-vt)$.  The total
energy density $\epsilon(x) = V(\phi) + \frac{1}{2}\phi_x^2$.  Thus,
the energy (or rest mass) of a kink is
\begin{eqnarray}
E_s &=& \int_{-\infty}^{\infty} \epsilon(x) dx \nonumber\\
&&= 4\xi
n\sqrt{n^2-\zeta^2} \left(\tanh^{-1}\sqrt{\frac{n-\zeta}
{n+\zeta}}\right) - 2\xi (n^2-\zeta^2).  \nonumber\\
\end{eqnarray}
The topological charge of the kink is 
\begin{eqnarray}
Q &=& \int_{-\infty}^{\infty} \frac{\partial\phi}{\partial x}dx =
\phi(+\infty)  - \phi(-\infty) \nonumber\\
&=& 2\tanh^{-1}\sqrt{\frac{n-\zeta}{n+\zeta}} = 2\phi_0~.  
\end{eqnarray}
The kinetic energy contribution of a moving kink with velocity $v$ is
obtained from $\phi\bigl((x-vt)/\xi\sqrt{1-v^2}\bigr)$ 
\begin{eqnarray}
&&E_{kin}= \int_{-\infty}^{\infty} dx
\frac{m}{2}\phi_t^2=\frac{m^{\ast}}{2}v^2~; \nonumber\\
&&m^{\ast}=\frac{m}{\xi\sqrt{1-v^2}}\left(\frac{2n}{n-\zeta}
\sqrt{\frac{n+\zeta}{n-\zeta}}\tanh^{-1}
\sqrt{\frac{n-\zeta}{n+\zeta}}-1\right)~.\nonumber\\ 
\end{eqnarray}
To calculate the contribution to the internal energy and specific heat 
from kinks we treat them as a one-dimensional gas of indistinguishable, 
independent, static particles of width $\sim2\xi$. The free energy 
contribution is then given by   
\begin{eqnarray}
&&F_K=\nonumber\\
&&-Nk_BT\frac{l}{\xi}\left(1+\frac{1}{2} \ln\frac{2\pi
k_BT}{m^{\ast}B^2}\right)\exp\left(-\frac{E_s}{k_BT
\sqrt{1-v^2}}\right)~, \nonumber\\
\end{eqnarray}
where $B$ is a phase space normalization constant. 
   
In order to understand kink-antikink interactions, it is useful to 
construct a kink lattice solution which is obtained by twice
integrating the equation of motion 
$$
\pm\sqrt \frac{2}{g} (x-x_0) = \int_{\phi(x_0)}^{\phi(x)} \frac{d\phi}
{\sqrt{V(\phi) - V_0}}~,  
$$
with the boundary conditions for a finite length $L = x_2 - x_1$: 
$\phi_x(x_1) = \phi_x(x_2) = 0$, and $V(\phi(x_1)) = V(\phi(x_2)) = 
V_0$.  Here $V_0$ and $x_0$ are integration constants and $V(x)>V_0$
for $x \in (x_1, x_2)$. The solution is 
\begin{eqnarray}
\phi_L(x) &=& \pm\tanh^{-1}\left(\tanh\phi_1~sn\left(\frac{x-x_0}
{\xi_L},k\right)\right)~; \nonumber\\
k &=& \frac{\tanh\phi_1}{\tanh\phi_2}~; ~~~d = 4K\xi_L;\nonumber\\
\xi_L &=& \frac{k}{2\zeta
\sinh\phi_1 \cosh\phi_2}\sqrt{\frac{g}{2}}
\end{eqnarray}
where $d$ is the periodicity of the kink lattice, $K(k)$ is the
complete elliptic integral of the first kind with modulus $k$,
$sn(x,k)$ is the Jacobi elliptic function, and  
\begin{eqnarray}
\cosh2\phi_{1,2} &=& \frac{n}{\zeta} \mp\frac{\sqrt{V_0}}{\zeta}~; ~~~k^2
= \frac{n^2-(\sqrt{V_0}+\zeta)^2} {n^2-(\sqrt{V_0}-\zeta)^2}~; \nonumber\\
&&0 < V_0 < V(\phi=0)=(\zeta-n)^2~.  
\end{eqnarray}
Equivalently, 
\begin{eqnarray}
&&\tanh^2\phi_{1,2} = \nonumber\\
&&\frac{n-\zeta}{n+\zeta}\left[\frac{(1+k^2)\zeta\mp
k'^2 (n+\zeta) + \sqrt{k'^4n^2+4k^2\zeta^2}}{(1+k^2)\zeta\mp
k'^2(n-\zeta) + \sqrt{k'^4n^2+4k^2\zeta^2}}\right]   
\end{eqnarray} 
and the characteristic length of a kink in the lattice  
\begin{eqnarray}
&&\xi_L^2=\frac{g}{2}\Big\{-n^2\nonumber\\
&&\zeta^2
\left[\frac{(1+k^2)\zeta^2-k'^2(n^2 -\zeta^2)+\zeta
\sqrt{k'^4n^2+4k^2\zeta^2}}{(1+k^2)\zeta^2+\zeta\sqrt{k'^4n^2  
+4k^2\zeta^2}}\right]^2\Big\}^{-1}  \nonumber\\
\end{eqnarray}
where $k'=\sqrt{1-k^2}$ is the complementary modulus. 

The topological charge (per period) in the lattice problem $Q_L =
2\phi_L(K) = 2\phi_1$ is smaller than the single kink case $Q=2 
\phi_0=2\tanh^{-1}\sqrt{(n-\zeta)/(n+\zeta)}$. Similarly, the kink
size in the kink lattice, $\xi_L$, is also smaller than the free kink
size $\xi$.

The energy of the kink lattice per period ({\em i.e.}, the energy per 
kink-antikink pair plus the interaction energy) is obtained, after
considerable algebra, using  $\phi_L(x)$ in Eq. (1) and evaluating
various integrals involving Jacobi elliptic functions \cite{byrd}:    
$$ 
E_L = 4\xi_L\left[(n+\zeta)^2K+\frac{\xi^2}{\xi_L^2}(n^2-\zeta^2)(K-E) 
-4n\zeta\Pi\right]~,  
$$ 
where $E(k)$ and $\Pi(\tanh^2\phi_1,k)$ are complete elliptic integrals
of second and third kind, respectively. Note that in the dilute limit
($k \rightarrow 1$, $d \rightarrow \infty$) the divergences in $K(k)$
and $\Pi(\tanh^2\phi_1,k)$ exactly cancel out and we recover the
single kink result $E_s$. The interaction energy as a function of
separation ({\em i.e.}, $k$ or $d$) is given by $E_{in} = E_L-2E_s$.   

The repulsive (attractive) kink-kink (kink-antikink) interaction in the 
asymptotic limit can be calculated by letting $k\rightarrow 1$ or $ k' 
\rightarrow 0$ and expanding $E_L$ up to order $k'^{4}$.  However, for 
the DSHG model it can directly be obtained from the asymptotic form of 
the single kink solution by using Manton's formula \cite{manton}, 
$$ 
U(r) =\pm\frac{2}{\xi}\frac{n^2-\zeta^2}{\zeta^2}
\exp\left(-{d\over\xi}\right) . 
$$ 

\section{Kink Lattice Thermodynamics} 
 
One can compute various ``thermodynamic'' quantities associated with
the kink lattice similar to the sine-Gordon case \cite{guptas}.  The 
internal energy per kink $U/N = E_L/2$ and $L/N=d$ imply that the 
thermodynamic pressure 
\begin{eqnarray}
&&P = -\left(\frac{\partial U}{\partial L}\right)_{T=0} =\nonumber\\
&&\frac{1}{2}
\left[2n\frac{\xi}{\xi_L}(n^2-\zeta^2)^{1/2}E(\beta,k) -
\frac{\xi^2}{\xi_L^2}(n^2 -\zeta^2) - (n-\zeta)^2\right] \nonumber\\ 
\end{eqnarray}
where $\beta = \sin^{-1}\tanh\phi_2$. The enthalpy is $H=(U+PL) =
N\mu$, with the chemical potential given by   
$$ 
\mu =
2E(n^2-\xi^2)\frac{\xi^2}{\xi_L^2}
\left[\frac{2n\xi_L}{\xi(n^2-\zeta)^{1/2}}F(\beta,k)-1\right]~, 
$$ 
where $F(\beta,k)$ and $E(\beta,k)$ are incomplete elliptic integrals
of the first and second kind, respectively.  Similarly, the isothermal
compressibility $H_T=-(1/L)(\partial L/\partial P)_T$ can be
calculated:   
\begin{eqnarray}
H_T &=& {1\over 2}\left\{1+[\xi_L/k(\partial\xi_L/\partial
k)][E/k'^2K-1]\right\} \nonumber\\
&&\times\Big\{(n\xi/\xi_L)\sqrt{n^2-\zeta^2}E(\beta,k)-(n^2-\zeta^2)
(\xi^2/\xi_L^2) \nonumber\\
&&-(n\xi/k)\sqrt{n^2-\zeta^2}(\partial\beta/\partial
k)(\partial\xi_L/\partial
k)\nonumber\\
&&\times\{E(\beta,k)-F(\beta,k)\}\Big\}^{-1}
\end{eqnarray}
where $\partial\beta/\partial k=\cosh\phi_2(\partial
\tanh\phi_2/\partial k)$ and $\partial\xi_L/\partial k$ can be
obtained from the above expressions for $\tanh\phi_2$ and $\xi_L$ as a
function of $k$.   

\section{Stability of kinks} 

Since the transformation $\phi\rightarrow \tanh\phi$ connects the
equation of motion (see below) for the $\phi^4$ and DSHG models, the
linear stability of the kink and kink lattice solutions in the two
models is directly related.  We find that the DSHG kink stability
equation is the Heun's equation which has at least two bound states:
(1) the usual zero frequency Goldstone mode and (2) a kink shape
oscillation mode.  Similarly, for linear stability 
of the kink lattice we can write $\tanh(\phi_L,t)=\tanh(\phi_L) +
\phi_L(x,t)$. Assuming a harmonic time variation
$\phi_L(x,t)=\psi(x)\exp(i\omega t)$ we find that the kink lattice
stability equation is also Heun's equation. A special case is the
Lam\'e equation of order two ($\nu=2$) which arises in the context of
the stability of the $\phi^4$ model kink lattice solution:  
$$
\psi_{xx}+[A(\omega,k)-\nu(1+\nu)k^2 sn^2x]\psi=0~. 
$$ The solutions of this equation are the Lam\'e functions
\cite{horov} with $2\nu+1$ eigenvalues. This implies a maximum of five
bound states.  Note that in the integrable case of the sine-Gordon
equation one gets the Lam\'e equation of order one ($\nu=1$) implying
a maximum of three bound states for the kink lattice and only the
Goldstone mode for the kink \cite{guptas}. However, the DSHG Heun's
equation does not reduce to a Lam\'e equation. These results apply to
linear stability (kink-phonon scattering) and cannot be readily
generalized to the case of nonlinear stability.
 
\section{Phonons}

The equation of motion for the field is given by
$$
m\phi_{tt} - \phi'' + 2\zeta[\zeta \sinh4\phi-2n \sinh2\phi]=0~. 
$$ 
This equation can be linearized around (1) $\phi = 0$ or (2) $\phi =
\phi_{0}$ leading to higher and lower energy phonons, respectively. 
For phonons around $\phi = 0$ the above equation can be approximated
by 
$$ 
m\phi_{tt} - \phi'' +
(8\zeta^2-8n\zeta)\phi+\frac{16}{3}(4\zeta^2-n\zeta)  
\phi^3=0~.
$$ 
For small amplitude oscillations ($\phi \ll 1$) we retain only the
linear term.  Assuming $\phi(x,t) = \overline\phi \exp[i(qx-\omega_q
t)]$ we get the dispersion relation
$$\omega_q^2=q^2-8\zeta(n-\zeta)~, $$ 
where the frequency $\omega_q$ will be real only for finite $q^2 \geq
8\zeta(n-\zeta)$. Of course, for $n < \zeta $ (when $\phi = 0$ is the
minium of the potential), the frequency $\omega_q$ is always real.

Similarly, the phonon dispersion around the minima $\pm\phi_0$ for
this model is  
$$\omega_q^2=q^2+8(n^2-\zeta^2)~.$$ 
Note that the frequencies are real for all $q \geq 0$ (and $n >
\zeta$) and the quantity $4(n^2-\zeta^2)$ is equivalent to the
coefficient $|A|$ of the quadratic term in the corresponding $\phi^4$
model \cite{krum}. The associated phonon contribution to the free
energy is   
\begin{equation}
F_{vib}={1\over 2\pi\delta}\ln\left({2\pi\over\delta\beta}\right)
+{1\over\beta}\sqrt{2(n^2-\zeta^2)}~,
\end{equation}
with $\delta$ being the lattice constant and $\beta\equiv 1/k_BT$.
The contribution to the internal energy and specific heat from 
phonons and kinks can be directly calculated using $F_{vib}$ and $F_K$, 
respectively.  However, here we do not consider the contribution to 
the free energy due to kink-phonon interactions \cite{currie}.  

\section{Transfer operator formalism}

The transfer operator method transforms the problem of finding the
canonical partition function for a system to an exactly equivalent
problem of finding the eigenvalues of a certain integral operator.  In
the continuum limit, this problem can be further reduced to finding
the energy eigenvalues of a related quantum mechanical problem ({\em
i.e.}, a Schr\"odinger equation with an effective potential). In
addition, the transfer operator formalism also provides exact
expressions for correlation functions.  The key property of the
DSHG potential (\ref{1a}) is that the $n$ lowest eigenvalues
and wavefunctions are known exactly, in contrast to other QES
systems. This enables exact evaluation of the PDF and the correlation
functions $C_1$ and $C_2$ (see below) at several temperatures. Aside
from validation of Langevin results against these quantities, the
exact single kink solution can be checked against that obtained from
the low temperature Langevin field configuration.

Turning now to the computation of $Z_{cl}$, we note that this
calculation can be divided into two parts: a trivial Gaussian
integration over the field momentum, and a computation of the
configurational partition function, which via the transfer integral
method becomes equivalent to solving a Schr\"odinger-like equation
\cite{ssf}. The Hamiltonian for the DSHG theory is
\begin{equation}
H=\int dx\left[{1\over 2}\pi^2+{1\over
2}\phi_x^2+V_{DSHG}(\phi)\right]
\end{equation}
and this leads to the Schr\"odinger-like equation for the eigenvalues
and eigenfunctions of the transfer operator,
\begin{equation}
-{1\over 2\beta^2}{\partial^2\over\partial\phi^2}\Psi_k
+(\zeta\cosh 2\phi-n)^2\Psi_k=E_k\Psi_k~.
\label{seqn}
\end{equation}
This is a QES system for which, at $2\beta^2 = 1$, using results for
a related potential from Ref. \cite{razavy}, the eigenstates of the
first $n$ levels can be found for $n = 1,2,3,4$. (We have been able to
extend this to the cases $n = 5,6$.) However, what one really wants is
to consider a given fixed-$n$ theory and obtain eigenstates at {\em
different} temperatures. It is easy to see from Eq. (\ref{seqn}), by
simple rescaling, that solutions of a fixed-$n$ theory at certain
values of $\beta$ are the same as the solutions of {\em another}
theory (different $n$ and $\zeta$) at $2\beta^2 = 1$. Depending on the
chosen value of $n$, exact solutions are available at different fixed
values of $\beta$. Here, we restrict ourselves to one such family
$(n=2)$ which allows the exact computation of the first few
eigenstates at $8\beta^2 = m^2~(m=1,\cdots,6)$. For illustration,
three examples of the (unnormalized) ground states are given below
(see also Fig. 2). The first (high temperature, $\beta^2=1/8$) has an
eigenfunction with a single peak while the other two (lower
temperatures, $\beta^2=1/2$ and $9/8$) have a double peak: 
\begin{eqnarray}
\Psi_0(\phi)|_{\beta^2={1\over 8}} & = & \exp\left(-{1\over 4}\zeta
\cosh 2\phi\right)~,\nonumber\\
\Psi_0(\phi)|_{\beta^2={1\over 2}} & = & \cosh\phi\exp\left(-{1\over
2} 
\zeta \cosh 2\phi\right)~,\nonumber\\
\Psi_0(\phi)|_{\beta^2={9\over 8}} & = & \left[3\zeta +
\left(1+\sqrt{1+9\zeta^2}\right)\cosh 2\phi\right] \nonumber\\
&&\times\exp\left(-{3\over
4} \zeta \cosh 2\phi\right)~,
\label{exactgs}
\end{eqnarray}
with corresponding ground state energies, $E_0=4+\zeta^2$,
$E_0=\zeta^2-2\zeta+3$ and $E_0 = \zeta^2 +{28\over 9} -{8 \over 9}
\sqrt{1+9\zeta^2}$ respectively. The PDF for the classical field
$\phi$ is just the square of the normalized ground state
eigenfunctions. Solutions at higher energies and other values of
$\beta$ are given in Appendix A.

Once the eigenvalues of the transfer operator are known, they can be
used to compute the correlation functions
$C_1(x)=\langle\phi(0)\phi(x)\rangle$ and
$C_2=\langle\delta\phi^2(0)\delta\phi^2(x)\rangle$, using
\begin{eqnarray}
C_1(x)&=&\sum_k |\langle\Psi_k|\phi|\Psi_0\rangle|^2
\exp\left[-\beta |x|(E_k-E_0)\right]~,
\label{c1}\\
C_2(x)&=&\sum_k |\langle\Psi_k|\delta\phi^2|\Psi_0\rangle |^2
\exp\left[-\beta |x|(E_k-E_0)\right]~.  \label{c2}
\end{eqnarray}
It is apparent that at large distances, $C_1$ and $C_2$ are dominated
by the lowest state with nonvanishing matrix elements: the first
excited state in the case of $C_1$ and the second excited state in the
case of $C_2$. Since $E_0$, $E_1$, and $E_2$ are known at certain
temperatures, the large distance behavior of these correlation
functions can be found exactly and compared with the results from
simulations. Static structure factors may also be calculated in much
the same way.

%PUT FIGURE 2 HERE:
\vspace{.5cm}
\centerline{\epsfig{figure=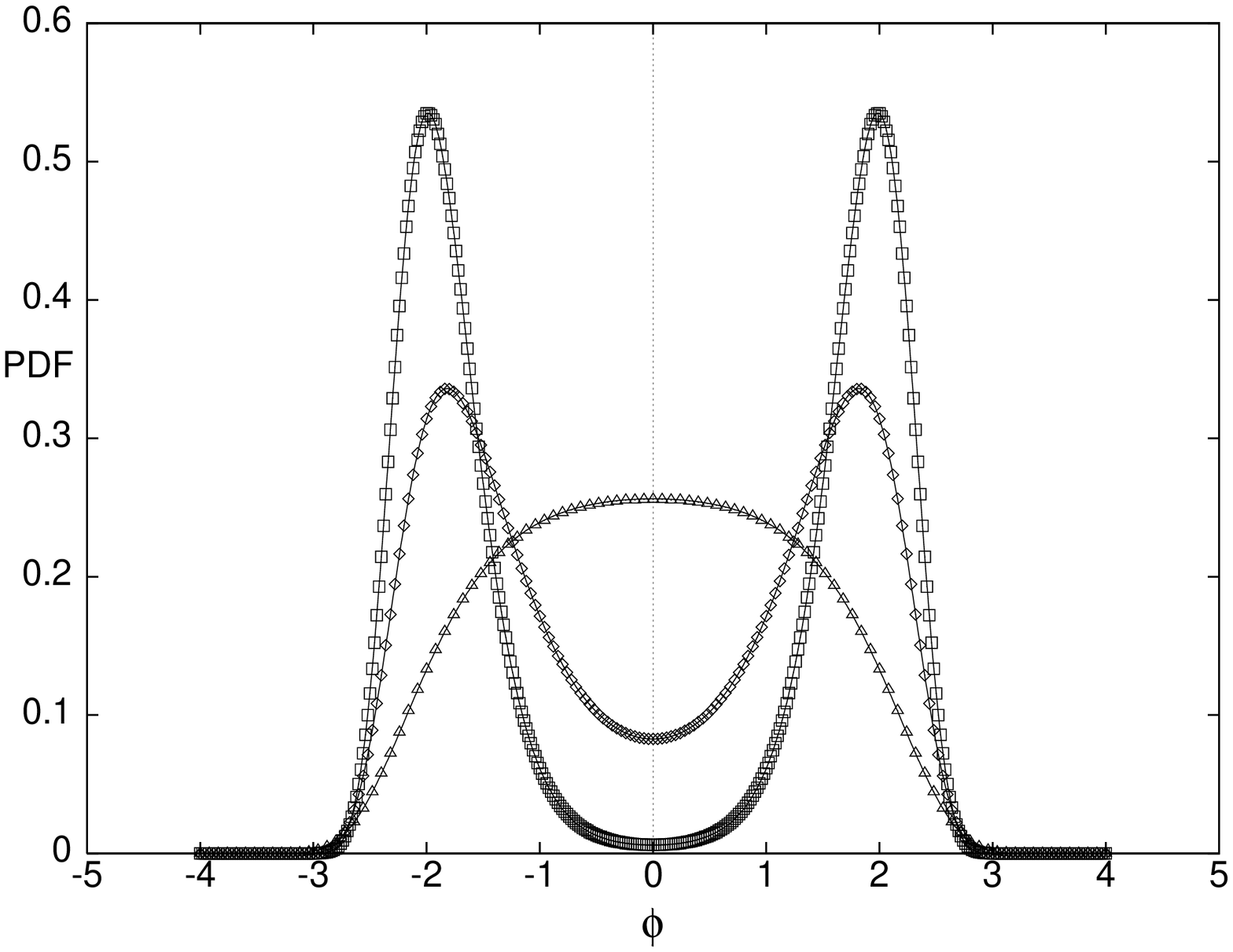,height=6cm,width=8cm,angle=0}}
\vspace{.25cm}
{FIG. 2. {\small{PDFs at three values of $\beta^2$ with the
continuum exact solutions shown as solid lines. Results from Langevin
simulations are superimposed at the $\beta$ values: 1/8 (triangles),
1/2 (diamonds), and 9/8 (squares).}}}
\vspace{.5cm}
 
At this point, it is important to mention the connection between the
``quantum'' calculations and kink physics. In the context of kink
statistical mechanics, it is usual to introduce a phenomenological
description of kinks as particles in a grand canonical
ensemble. However, this is unnecessary, and all such thermodynamical
information can be extracted directly from the Schr\"odinger
description of the transfer operator. For example, the kink density
has been obtained in this way in Ref. \cite{ahk}. Simpler quantities
like $C_1$ and $C_2$ have obvious natural interpretations in terms of
kinks. The $C_1$ correlation length is related to the kink/antikink
spacing and increases monotonically as $\beta$ increases (Fig. 3). The
behavior of the $C_2$ correlation length is more subtle, since $C_2$
is not sensitive to domain size. At both high temperatures (no kinks)
and low temperatures (number of kinks exponentially suppressed), the
correlation length is dominated by the thermal phonon
contribution. However, in the range of temperatures close to the kink 
transition the fluctuations on the kink length scale become important
and can dominate $C_2$. At these temperatures one might expect a
maximum in the $C_2$ correlation length, on the order of the kink
size, and this is indeed what we observe numerically (Fig. 4). The
Schottky anomaly in the specific heat \cite{ahk} arises for the very
same reason. 

%PUT FIGURE 3 HERE:
\vspace{.5cm}
\centerline{\epsfig{figure=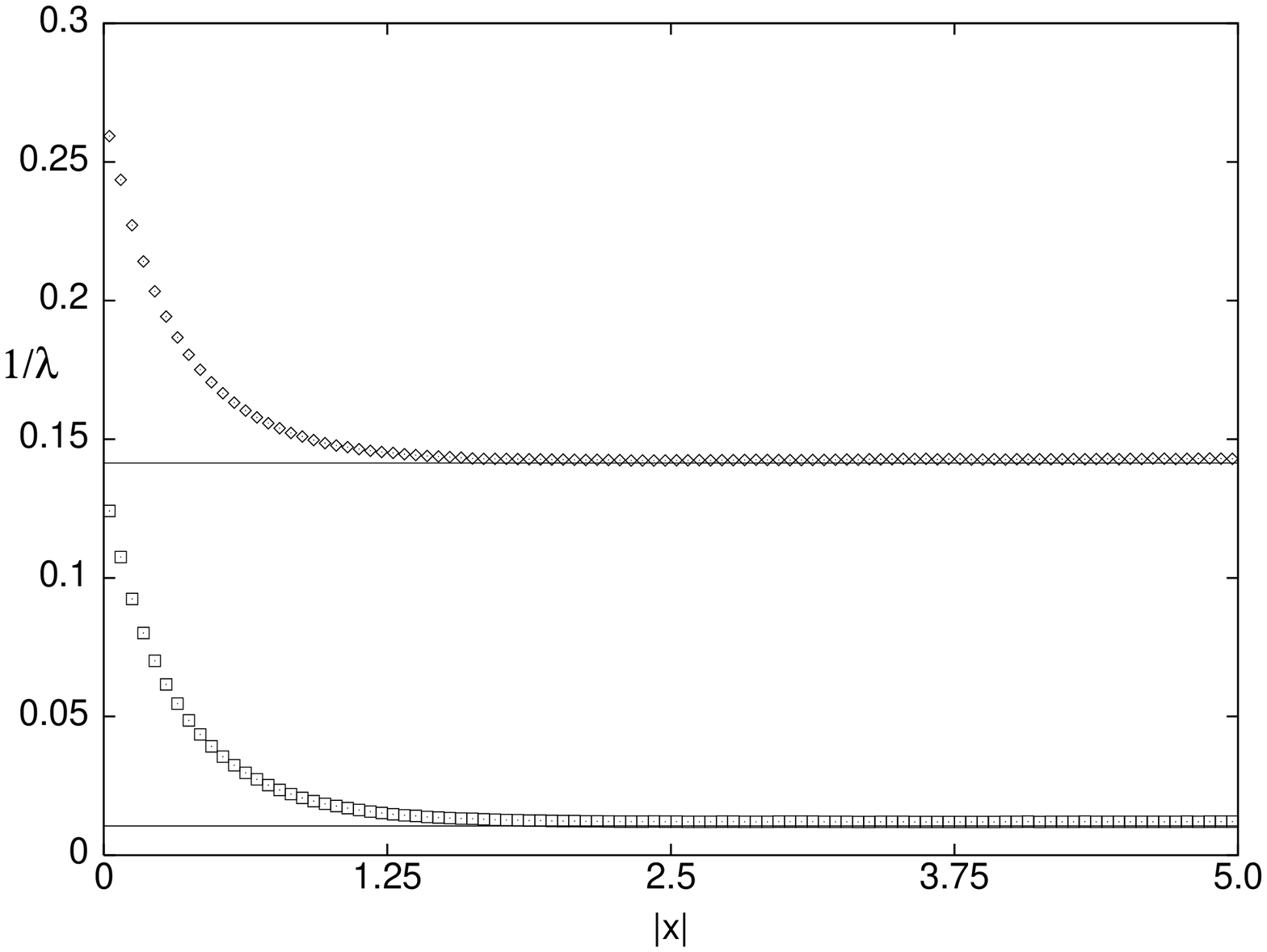,height=6cm,width=8cm,angle=0}}
\vspace{.25cm}
{FIG. 3. {\small{Comparison of the inverse correlation
lengths from $C_1(x)$ obtained via Langevin simulations [$\beta^2$:
1/2 (diamonds), and 9/8 (squares)] and from the large $|x|$ continuum
exact results (solid lines).}}}

%PUT FIGURE 4 HERE:
\vspace{.5cm}
\centerline{\epsfig{figure=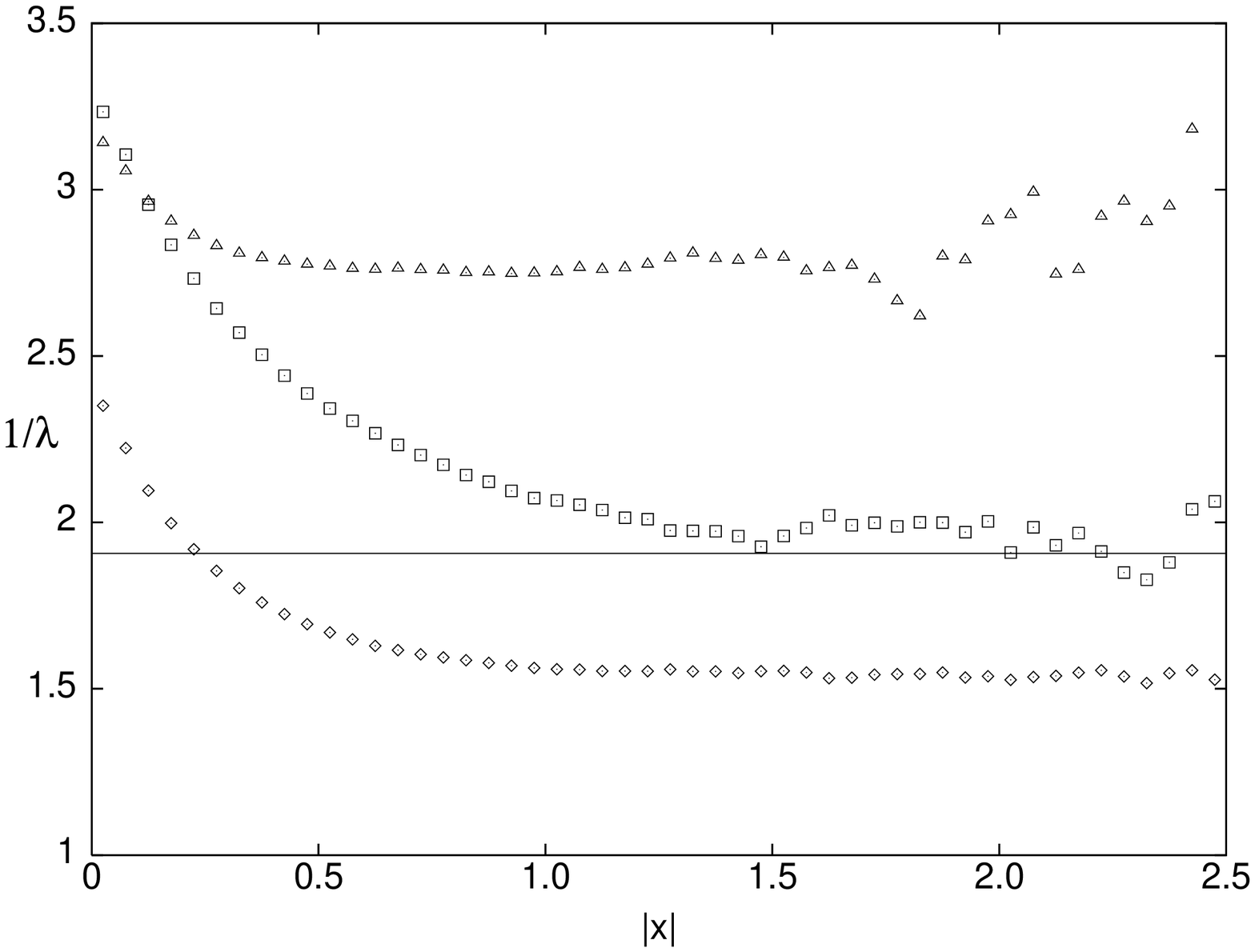,height=6cm,width=8cm,angle=0}}
\vspace{.25cm}
{FIG. 4. {\small{The numerical and exact inverse
correlation lengths from $C_2(x)$ are compared for the same three
temperatures and with the same conventions as in Fig. 2. The solid
line is the large $|x|$ continuum exact result for $\beta^2=9/8$. The
largest correlation length is at the intermediate value of $\beta$
(see text).}}}
 
\section{QES solutions}
The explicit QES solutions (exact eigenvalues and eigenfunctions) for
$2\beta^2 =1$ and $n=1,2,3,4$ are given by Razavy \cite{razavy} while
we have been able to extend these results to $n = 5,6$. We now explain
in more detail how to relate these results to a theory with a given
value of $n$ (say $n=2$) but with different values of $\beta$.  We
start from the Schr\"odinger equation (\ref{seqn}) at $2\beta^2=1$ and
arbitrary $n~(=1,2,\cdots,6)$. On multiplying both sides of this
equation by $4/n^2$, it immediately follows that the exact solutions
for a given value of $n, \zeta, E$ at $2\beta^2 = 1$ are equivalent to
the solution of the DSHG equation at $n=2$ but at $8\beta^2 =n^2$,
$\hat{E} = 4E/n^2,~\hat{\zeta} = 2\zeta/n$.  Clearly,
instead of fixing $n = 2$, one could choose any of the six values of
$n$ and for each case one knows the exact solutions for that theory at
six different temperatures. As an illustration, the explicit QES
solutions (exact eigenvalues and eigenfunctions) for $n = 1,2$ at six
different temperatures are presented in Appendix A.

\section{Connection with $\phi^4$ and double sine-Gordon models}

As a final point, we consider the relationship of the DSHG theory to
the more familiar Landau-Ginzburg model. Scrutiny of the static
equation of motion [$\phi_x^2=V(\phi)$] reveals an important
connection between the kink (and kink lattice) solutions of the
$\phi^4$ model and the double sine-Gordon (DSG) and DSHG
models. Consider the $\phi^4$ potential $V_4(u)
=\left[(n+\zeta)u^2-(n-\zeta)\right]^2$.  The substitution
$u=\tanh\phi$ takes the (static) equations of motion over to the DSHG
equations. The alternative substitution $u= \tan\phi$ leads to a
DSG model. This means that {\em all} known solutions of the $\phi^4$
equation of motion can be directly taken over to the DSHG and DSG
equations of motion (and vice versa). As one use of this interesting
relationship, the DSG kink lattice solution (not known heretofore in
the literature) can be written down directly in case $V_{DSG} =
(\zeta\cos 2\phi-n)^2$: 
\begin{equation}
\phi_L=\pm\tan^{-1}\left(\tan\phi_1~\hbox{sn}
\left({x-x_0\over\xi_L},k\right)\right)~,
\end{equation}
simply by using the substitution $\tanh\phi\rightarrow \tan\phi$ in
the kink lattice solution $\phi_L$ of the DSHG model.
 
This connection enables us to write down by inspection not just the
kink solutions but their total energy as well, which is often a very
tedious task. Similarly, knowing the linear stability of the $\phi^4$
kink and kink lattice, important results follow for the stability of
corresponding solutions of the DSG and DSHG models (see Sec. V).
Moreover, since we know that the DSHG model is an example of a QES
system, and considering the very similar way in which the DSG and DSHG
models are related to $\phi^4$, it is logical to conjecture that some
DSG model may also be a QES system. Indeed, this is the case, and we
have found several exact eigenvalues and eigenfunctions at many
temperatures for a particular DSG system. The exact statistical
mechanical results for that model, similar to the DSHG results
presented here, will be reported elsewhere \cite{hks}.

Alternatively, if we start from the $\phi^4$ potential $V_4(u) =
\left[(n+\zeta)u^2+(n-\zeta)\right]^2$, the transformation
$u=\tan\phi$ takes the (static) equations of motion over to the model
$V_{DSG} = [\zeta\cos 2\phi-n]^2$.  For $n < \zeta$ we get the doubly
periodic DSG potential with two types of kinks (``small'' and
``large'').  When $n > \zeta$, we get the singly periodic DSG
potential with the $2\pi$-kink solution (see Fig.~1 of
Ref. \cite{doubleSG}). The associated $V_4(u)$ in this case is a
single well potential.  However, the real transformation does not lead
to a QES DSG system unlike the transformation above. 

We note that the transformations $u= \tan\phi$ and $u=\tanh\phi$
connect the equation of motion of the $\phi^4$ model to that of the
DSG and DSHG systems, respectively.  On the other hand, the
transformations $u= \cos \phi$ and $u=\cosh\phi$ connect the equation
of motion of the $\phi^4$ model to that of the exactly solvable
sine-Gordon ($V_{SG}=\sin^2\phi$) and sinh-Gordon
($V_{SHG}=\sinh^2\phi$) systems, respectively, provided we start with
$V_4(u) = (u^2-1)^2$.  We also note that the number of periodicity in
the trigonometric potentials (DG, DSG, triple sine-Gordon (TSG),
$\cdots$) corresponds to the number of wells in the associated
hyperbolic potentials (SHG, DSHG, triple sinh-Gordon (TSHG), $\cdots$).
We conjecture that the equations of motion for the TSG and TSHG models
can be directly connected with that of the $\phi^6$ model via
appropriate transformations.  In addition, while the TSG equation
\cite{bullough,tsg} is known to arise in nonlinear optics the TSHG
potential [e.g. $V_{TSHG}=A\cosh2\phi+B\cosh4\phi+C\cosh6\phi$] can
serve as a model system to study first-order phase transitions.

\section{Simulation Results}

The numerical study of kink statistical mechanics was carried out via
Langevin simulations. The advantage of such simulations is that
representative field configurations (see Fig. 5) are available and can
be analyzed to compute such quantities as the kink density, finite
temperature kink profile, the spatial kink distribution, kink
transport and nucleation, {\em etc}. This sort of analysis is not
possible with Monte Carlo techniques.

%PUT FIGURE 5 HERE:
\vspace{.5cm}
\centerline{\epsfig{figure=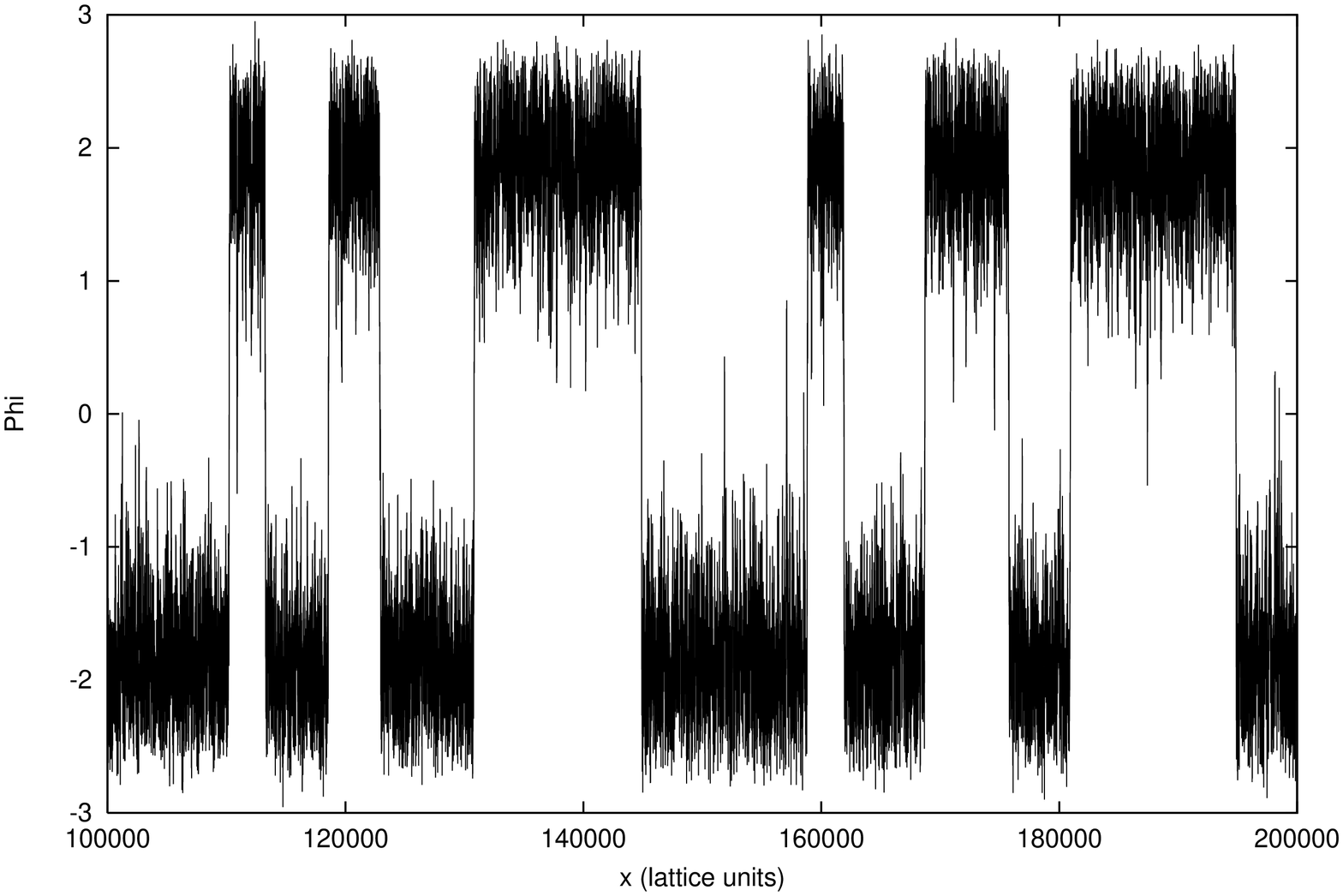,height=5cm,width=8cm,angle=-90}}
\vspace{.25cm}
{FIG. 5. {\small{Sample kink configurations from a $100K$
section of a $400K$ unit simulation with $\beta^2=9/8$.}}}

\vspace{.5cm}

The additive noise Langevin equation for the double sinh-Gordon model
is 
$$
\partial_{tt}^2\phi=\partial_{xx}^2\phi-\eta\partial_t\phi-
4\zeta(\zeta\cosh 2\phi-n)\sinh 2\phi+{\hat F}(x,t)
$$
where the viscosity $\eta$ and the Gaussian white noise ${\hat F}$ are
related by the fluctuation-dissipation theorem:
$$
\left\langle{\hat F}(x,t){\hat F}(x^{\prime},t^{\prime})\right\rangle=
2\eta\beta^{-1}\delta(x-x^{\prime}) \delta(t-t^{\prime})~.
$$
The lattice versions of the above continuous equations (with periodic
boundary conditions) were solved using standard techniques
\cite{LE}. Random initial conditions were driven to equilibrium and
the results sampled in time thereafter to yield time averaged PDFs,
{\em etc.} The use of the Langevin technique for obtaining
thermodynamic quantities is straightforward and as shown below, can be
remarkably accurate. The availability of exact solutions is very
useful since they set stringent criteria for accuracy that must be met
by numerical methods. In the simulations reported below we used a
lattice size of typically $512K$ points with a lattice spacing
$\Delta=0.025$. The time step was taken to be $\epsilon=0.005$. [We
employed both $\sqrt{\epsilon}$ (Euler) and $\epsilon^2$ (Runge-Kutta)
order stochastic integration algorithms.] These values 
are substantially smaller than what has been the norm so far in
Langevin simulations of other $1+1$ dimensional field theories
\cite{fjash}. In fact, based on our experience with the double
sinh-Gordon system it is likely that errors in previous Langevin
analyses have been as high as $30\%$.

Fig. 2 shows the striking agreement between the numerically obtained
and the exact continuum PDFs at three temperatures: The worst case
departure is at the level of parts per thousand. The comparisons for
the inverse correlation lengths are given in Figs. 3 and 4. DSHG
system parameters are $n=2$, $\zeta=0.05$. For $C_1$, the numerical
values are $1/\lambda=0.1425$ ($\beta^2=1/2$) and $1/\lambda=0.012$
($\beta^2=9/8$) as compared to the exact values in the continuum
theory of $0.14142$ and $.0105$, respectively. The small offset
between the continuum and lattice calculations is due to the finite
value of the lattice constant and is consistent with estimates from
higher-order contributions to the transfer integral \cite{shgl}.
The kink number density at low temperatures is related to the
correlation length via $N_k\simeq 1/4\lambda$ \cite{ahk}. This
general relationship is borne out in the DSHG simulations.

The high quality of these numerical simulations implies that the PDF
can now be used directly to compute thermodynamic quantities at {\em
any} temperature. Since the PDF is just the square of the ground state
wave function of the Schr\"odinger equation, one can use it to compute
the ground state energy $E_0$ numerically, from which the internal
energy ($U=\partial E_0/\partial \beta$), the free energy
($F=E_0/\beta$), and the entropy ($S=\beta \partial
E_0/\partial\beta-E_0$) can all be computed in a straightforward
way. The specific heat involves two $\beta$ derivatives and is
difficult to obtain with good accuracy but in this case, the standard
energy fluctuation method is quite effective. The use of the PDF
complements traditional techniques utilizing energy fluctuations in
Langevin simulations which are not suited to free energy and entropy
calculations.

The QES nature of the DSHG theory allows not only the exact
computation of $E_0$ at several temperatures, but also of $\partial
E_0/\partial \beta$, using first order perturbation theory: $\partial
E_0/\partial \beta |_{\beta=\beta_0}=
(\Psi_0,\partial^2\Psi_0/\partial\phi^2)$ where $\beta_0$ is one of
the special checkpoint temperatures. Thus the internal energy $U$ and
the entropy $S$ can also be found exactly at these temperatures. Once
again, these quantities can be used to validate numerical work over a
broad range of temperatures.

\section{Conclusion}

In this paper we have analytically obtained the free energy of the
kink-bearing DSHG theory in $1+1$ dimensions at several
temperatures. To our knowledge, this is the first such analytical
calculation at several temperatures. We have compared the analytical
results against large scale Langevin simulations and found excellent
agreement between the two. In view of this agreement, the study of
PDF-based thermodynamics via Langevin simulations appears quite
tractable. It is worth emphasizing that the calibration of Langevin
simulations constitutes a fundamental problem in the numerical study
of nonlinear dynamical systems. Numerical convergence of results is
certainly an important test that does not require the knowledge of an
exact solution. However, standard finite difference error analysis
cannot be applied directly to Langevin simulations because of their
stochastic nature. Moreover, the error and convergence analysis for
nonlinear stochastic PDEs with spatio-temporal noise remains to be
completely worked out. For this reason, a nontrivial test system for
which analytic solutions are available at several temperatures
provides a valuable tool for testing the Langevin code. In fact based
on our experience in the DSHG case, it is likely that previous
Langevin-based analyses have errors as high as 30\%. We believe that
apart from the intrinsic physical interest of the the double-well DSHG
model, this system has the potential of becoming a standard benchmark
problem in numerical simulations of Langevin systems.
  
We have also analytically studied the static equations of motion of
the DSHG system and found an interesting connection between the
$\phi^4$-theory, DSHG theory and a DSG theory. As a result, the kink and
lattice kink solutions as well as kink total energy can be directly
written down for any one of them in terms of the other. This is
interesting as the lattice kink solution of this DSG case, which was
unknown so far, can be immediately written down.  Further, because of
the intimate connection between the DSHG and the DSG case, it is
reasonable to expect that a DSG theory may also be a QES system. We
have been able to obtain several eigenstates of the DSG case at
various temperatures from this identification \cite{hks}. It is
clearly of great interest to carry out high quality Langevin
simulations of the DSG case, which is a periodic system, and compare
the agreement between the analytical and the simulation results. We
hope to report on this problem in the near future.

\section{Acknowledgment}

This work was supported by the U.S. Department of Energy at Los Alamos
National Laboratory. AK thanks the Theoretical Division of Los Alamos
National Laboratory for hospitality. SH acknowledges useful
discussions with Grant Lythe. Numerical simulations were performed on
the CM-5 and the Origin 2000 at the Advanced Computing Laboratory, Los
Alamos National Laboratory, and on the T3E at the National Energy
Research Scientific Computing Center, Lawrence Berkeley National
Laboratory.

\appendix 
\section{QES solutions} 

We obtain two sets of exact solutions for the Schr\"odinger equation
(\ref{seqn}).\\

\noindent{\bf Set-I ($n=1$)}: Note that $\epsilon_0$, as it occurs in 
the free energy, is related to the ground state energy $E_0$ by 
$\epsilon_0 = E_0 -V_{min}$. For $n=1$, $V_{min} =0$ for $\zeta <1$
while $V_{min}=2\zeta -\zeta^2 -1$ for $\zeta >1$. However, in either
case, $\epsilon_n - \epsilon_m = E_n - E_m$.\\

(1) For $2\beta^2=1$: 
\begin{equation}
\Psi_0(\phi) = \exp\left[-\frac{\zeta}{2}\cosh2\phi\right]~, 
~ E_0 = 1+\zeta^2~. 
\end{equation}

(2) For $2\beta^2=4$: 
\begin{eqnarray}
\Psi_0(\phi) &=& \cosh\phi \exp[-{\zeta}\cosh2\phi]~,  
~E_0  = \zeta^2-\zeta+\frac{3}{4}~,\nonumber\\
\Psi_1(\phi) &=& \sinh\phi \exp[-\zeta \cosh2\phi]~,
~E_1 = \zeta^2+\zeta+\frac{3}{4}~.   \nonumber\\  
\end{eqnarray}
Note that $\epsilon_1-\epsilon_0 = E_1 - E_0 = 2\zeta$.\\

(3) For $2\beta^2=9$:
\begin{eqnarray}
\Psi_0(\phi) &=&
[6\zeta+(1+\sqrt{1+36\zeta^2})\cosh2\phi]\nonumber\\
&&\times\exp\left[-\frac{3\zeta}{2}
\cosh2\phi\right]~,\nonumber\\
E_0 &=& \zeta^2-\frac{2}{9}\sqrt{1+36\zeta^2}+\frac{7}{9}~,
\nonumber\\    
\Psi_1(\phi) &=& \sinh2\phi
\exp\left[-\frac{3\zeta}{2}\cosh2\phi\right]~,\nonumber\\
E_1&=&\zeta^2+\frac{5}{9}~,  \nonumber\\     
\Psi_2(\phi) &=&
[6\zeta-(\sqrt{1+36\zeta^2}-1)\cosh2\phi]\nonumber\\
&\times&\exp\left[-\frac{3\zeta}{2}  
\cosh2\phi\right]~,\nonumber\\
E_2 &=& \zeta^2+\frac{1}{9}(2\sqrt{1+36\zeta^2}+7)~.
\end{eqnarray}
Note that $\epsilon_1-\epsilon_0=\frac{2}{9}(\sqrt{1+36\zeta^2}-1)$
and  $\epsilon_2-\epsilon_0 = {4 \over 9} \sqrt{1+36\zeta^2}$.\\ 
  
(4) For $2\beta^2=16$: 
\begin{eqnarray}
\Psi_0(\phi) &=& [12\zeta
\cosh\phi+(2-4\zeta+2\sqrt{1-4\zeta+16\zeta^2})\nonumber\\
&\times&\cosh 3\phi]\exp[-2\zeta \cosh2\phi]~,   \nonumber\\
E_0&=&\frac{1}{16}(16\zeta^2-4\sqrt{1-4\zeta+16\zeta^2}-8\zeta+11)~,
\nonumber\\ 
\Psi_1(\phi) &=& [12\zeta
\sinh\phi+(2+4\zeta+2\sqrt{1+4\zeta+16\zeta^2})\nonumber\\
&\times&\sinh3\phi]\exp[-2\zeta \cosh2\phi]~,    \nonumber\\
E_1 &=&\frac{1}{16}(16\zeta^2+8\zeta-4\sqrt{1+4\zeta+16\zeta^2}+11)
~,  \nonumber\\
\Psi_2(\phi) &=&
[12\zeta\cosh\phi+(2-4\zeta-2\sqrt{1-4\zeta+16\zeta^2})\nonumber\\
&\times&\cosh 3\phi]\exp[-2\zeta \cosh2\phi]~,     \nonumber\\
E_2&=&\frac{1}{16}(16\zeta^2-8\zeta+4\sqrt{1-4\zeta+16\zeta^2}+11)~.
\nonumber\\  
\end{eqnarray}
Note that $\epsilon_1-\epsilon_0=\zeta+{1 \over
4}\sqrt{1-4\zeta+16\zeta^2} -{1 \over 4} 
\sqrt{1+4\zeta+16\zeta^2}$ and $\epsilon_2-\epsilon_0
= {1 \over 2}\sqrt{1-4\zeta 
+16\zeta^2}$. \\
 
\noindent{\bf Set-II ($n=2$)}: In this case $V_{min}$ for $\zeta>2$ is
$4\zeta -\zeta^2 -4$ while $V_{min}=0$ in case $\zeta<2$. \\

(1) For $2\beta^2=1$: 
\begin{eqnarray}
\Psi_0(\phi) &=& \cosh\phi
\exp\left[-\frac{\zeta}{2}\cosh2\phi\right]~,   
~~E_0 =\zeta^2-2\zeta+3 ~, \nonumber\\
\Psi_1(\phi) &=& \sinh\phi
\exp\left[-\frac{\zeta}{2}\cosh2\phi\right]~,   
~~E_1 =\zeta^2+2\zeta+3~. \nonumber\\ 
\end{eqnarray}
Note that $\epsilon_1 - \epsilon_0 = 4\zeta$.\\

(2) For $2\beta^2=4$: 
\begin{eqnarray} 
\Psi_0(\phi) &=& [3\zeta
\cosh\phi+(1-\zeta+\sqrt{1-2\zeta+4\zeta^2})\cosh 3\zeta]\nonumber\\
&\times&\exp[-\zeta\cosh2\phi]~, \nonumber\\  
E_0&=&\frac{1}{4}(4\zeta^2-4\sqrt{1-2\zeta+4\zeta^2}-4\zeta+11)~,
\nonumber\\    
\Psi_1(\phi) &=& [3\zeta
\sinh\phi+(1+\zeta+\sqrt{1+2\zeta+4\zeta^2})\sinh 3\phi]\nonumber\\
&\times&\exp[-\zeta\cosh2\phi]~, \nonumber\\
E_1&=&\frac{1}{4}(4\zeta^2+4\zeta-4\sqrt{1+2\zeta+4\zeta^2}+11)~,
\nonumber\\ 
\Psi_2(\phi) &=& [3\zeta
\cosh\phi+(1-\zeta-\sqrt{1-2\zeta+4\zeta^2})\cosh 3\phi]\nonumber\\
&\times&\exp[-\zeta\cosh2\phi]~,  \nonumber\\
E_2&=&\frac{1}{4}(4\zeta^2-4\zeta+4\sqrt{1-2\zeta+4\zeta^2}+11)~.    
\end{eqnarray}
Note that $\epsilon_1 - \epsilon_0 = 2\zeta +\sqrt{1-2\zeta
+4\zeta^2}-\sqrt{1+2\zeta +4\zeta^2}$ while $\epsilon_2 -\epsilon_0  
= 2 \sqrt{1-2\zeta +4\zeta^2}$.\\

(3) For $2\beta^2=1/4$:
\begin{equation}
\Psi_0(\phi) = \exp\left[-\frac{\zeta}{4}\cosh2\phi\right]~, 
~~E_0=\zeta^2+4~.  
\end{equation}

(4) For $2\beta^2=9/4$:
\begin{eqnarray}
\Psi_0(\phi) &=&
[3\zeta+(1+\sqrt{1+9\zeta^2})\cosh2\phi]\nonumber\\
&\times&\exp\left[-\frac{3\zeta}{4}\cosh2\phi\right]~, \nonumber\\  
E_0&=&\zeta^2-\frac{8}{9}\sqrt{1+9\zeta^2}+\frac{28}{9}~, \nonumber\\    
\Psi_1(\phi) &=& \sinh2\phi
\exp\left[-\frac{3\zeta}{4}\cosh2\phi\right]~,  \nonumber\\ 
E_1&=&\zeta^2+\frac{20}{9}~,  \nonumber\\
\Psi_2(\phi) &=& [3\zeta-(\sqrt{1+9\zeta^2}-1)\cosh2\phi]\nonumber\\
&\times&\exp\left[-\frac{3\zeta}{4} \cosh2\phi\right]~,\nonumber\\  
E_2 &=& \zeta^2+\frac{4}{9}\left(2\sqrt{1+9\zeta^2}+7\right)~. 
\end{eqnarray}
Note that $\epsilon_1-\epsilon_0=\frac{8}{9} [\sqrt{1+9\zeta^2} - 1]$
and $\epsilon_2-\epsilon_0=\frac{16}{9}\sqrt{1+9\zeta^2}$.  

We now show that the first five and six levels of the DSHG potential
as given by Eq. (\ref{1a}) can be written down exactly in case $n=5$
and $n=6$ respectively (at $2\beta^2 = 1$). Using the trivial scaling,
one can then obtain exact solutions for say $n=2$ but at $2\beta^2 =
25/4$ and $2\beta^2=9$.  For example, for $n=5$, $\hat{E}_{0,2,4}$ are
solutions of the cubic equation
\begin{equation}
\hat{E}(\hat{E}+4)(\hat{E}+16)-64\zeta^2 \hat{E}-768\zeta^2=0~,
\end{equation}
and the corresponding eigenfunctions are given by 
\begin{equation}
\psi(\phi) = v(\phi) \exp\left[-\frac{\zeta}{2}\cosh2\phi\right]~,
\end{equation}
where 
\begin{eqnarray}
v_{0,2,4}&=&6\zeta-\hat{E} \cosh 2\phi+\frac{2\zeta
\hat{E}}{\hat{E}+16} \cosh 4\phi~,\nonumber\\
\hat{E}_{1,3}&=&-10\mp2\sqrt{4\zeta^2+9}~,\nonumber\\
v_{1,3}&=&4\zeta \sinh 2\phi+(3\pm\sqrt{4\zeta^2+9}) \sinh
4\phi~.\nonumber\\ 
\end{eqnarray}
Note that here $\hat{E} = E -\zeta^2 - 25$.
 
For n=6, $\hat{E}_{0,2,4}$ are solutions of the cubic equation 
\begin{eqnarray}
&&(\hat{E}+9)(\hat{E}+25)(\hat{E}+6\zeta+1)-52\zeta^2\hat{E}\nonumber\\
&&-820\zeta^2-120\zeta^3=0~,
\end{eqnarray}
and 
\begin{eqnarray}
v_{0,2,4}&=&8\zeta \cosh\phi-(\hat{E}+6\zeta+1)\cosh
3\phi \nonumber\\
&&+\frac{2\zeta(\hat{E}+  
6\zeta+1)}{\hat{E}+25} \cosh 5\phi~.
\end{eqnarray}
Note that $\hat{E}_{1,3,5}$ are solutions of the same cubic equation
with $\zeta$ replaced by $-\zeta$, and 
\begin{eqnarray}
v_{1,3,5}&=&8\zeta \sinh \phi -(\hat{E}-6\zeta+1) \sinh 3\phi
\nonumber\\ 
&&+\frac{2\zeta(\hat{E}- 
6\zeta+1)}{\hat{E}+25} \sinh 5\phi~.
\end{eqnarray}
Here $\hat{E} = E -\zeta^2 - 36$.

%\narrowtext

\end{document}